\documentclass[openacc]{rsproca_new}

\begin{document}

\title{Introduction to the special issue dedicated to Michael J.~Duff FRS on the occasion of his 70$^{\rm th}$ birthday}

\author{
L. Borsten$^{1}$, A. Marrani$^{2}$, C. N.~Pope$^{3,4}$, K. Stelle$^{5}$}

\address{$^{1}$Maxwell Institute for Mathematical Sciences \&
Department of Mathematics, Heriot--Watt University, Edinburgh EH14 4AS, UK\\
$^{2}$Instituto de F\'isica Teorica, Departamento de F\'isica,
Universidad de Murcia, Campus de Espinardo, E-30100, Spain\\
$^{3}$George P. \& Cynthia W. Mitchell Institute for Fundamental Physics, Texas A\&M University, College Station, TX 77843-4242, USA\\
$^{4}$ DAMTP, Centre for Mathematical Sciences,
Cambridge University, Wilberforce Road, Cambridge CB3 OWA, UK\\
$^{5}$The Blackett Laboratory, Imperial College London Prince Consort Road, London SW7 2AZ}

\subject{Quantum gravity, branes, M-theory}


\corres{Leron Borsten\\
\email{l.borsten@herts.ac.uk}}

\begin{abstract}
Introduction to the special issue dedicated to Michael J.~Duff FRS on the occasion of his 70$^{\rm th}$ birthday, with a brief scientific biography.
\end{abstract}


\begin{fmtext}

\section{Michael J.~Duff: Quantum gravity, branes, M-theory}

This  special feature, dedicated to Michael J. Duff FRS on the occasion of his 70th birthday, concerns topics in `Quantum gravity, branes and M-theory'. These three intertwining subjects have been central to Duff's work; indeed many of his contributions have come to define  significant aspects of what we actually mean by  these terms. From the discovery of Weyl anomalies to  recognising  superstrings in 10 dimensions as a special case of membranes in an 11-dimensional M-theory, Duff's insights have shaped major developments across these themes. So it is an apposite setting for such a celebration and we are delighted to be able to include in 
\end{fmtext}


\maketitle

\noindent   this collection contributions from many of the pioneers of quantum gravity, branes and M-theory.  The breadth of these topics has placed little constraint on the multiplicity of ideas appearing in these pages, from astrophysical black holes to chaotic condensed matter. Again, this is fitting as Duff's scientific remit  spans a remarkable diversity of motifs, from the fundamentals of M-theory to entanglement in quantum information.

\section{A brief scientific biography of Michael J.~Duff FRS}
Michael J.~Duff FRS (Mike, from here onwards) did his PhD at Imperial College London under the supervision of  Nobel Laureate Abdus Salam KBE FRS, with mentorship also from Christopher J.~Isham\footnote{See \cite{Duff:2017zff, Duff:2021rqf} for  personal tributes to Salam and Isham, authored by Mike.}.  He  was somewhat thrown in at the deep end, charged with resolving a bet between Salam and Sir Hermann Bondi KCB FRS, Nobel Laureate Sir Roger Penrose OM FRS HonFInstP and John Archibald Wheeler, some of the most influential quantum field theorists and general relativists of the twentieth century.   Salam maintained that the Schwarzschild black hole solution of general relativity could be perturbatively  reconstructed via the Feynman diagrams of quantum field theory. Mike confirmed this speculation \cite{Duff:1973zz} in a calculation that could be regarded as an early precursor to a now thriving industry applying  scattering amplitudes to classical general relativity \cite{Bern:2021yeh}\footnote{Before we go any further, we should first sincerely apologise for our  omission of the many crucial contributions by others to the story we shall tell.  Giving due credit even only to  those with direct connections to Mike, with anything like an even hand, would turn this lightning summary into a full blown review of the development of M-theory.}. Mike  took up his first postdoctoral position at the International Centre for Theoretical Physics (ICTP), Trieste, Italy, recently established by Salam and so the destination of choice for many a prot\'eg\'e. There, in a follow-up paper,  Mike showed that loop contributions  implied a $1/r^3$ correction to the classical Schwarzschild  solution. One should keep in mind  that  the problem of quantum gravity was still viewed with  suspicion, or even contempt\footnote{Mike recalls  his thesis title ``Problems in the Classical and Quantum Theories of Gravitation'' being met with ``hoots of derision'' \cite{Duff:1993wm}. How things have changed.}, in certain quarters.   Mike would then initiate a fruitful collaboration with  Derek M.~Capper, who had also recently taken the road from Imperial to ICTP, and Leopold Halpern, further developing the interface between quantum theory and  gravity \cite{Capper:1973mv, Capper:1974ed}. Although this early work was somewhat forgotten for a period,  it pre-empted many future and current themes in  quantum gravity. As we shall see, the farsightedness of Mike's  work would become a recurring theme. 

 On returning to the UK as part of  Dennis Sciama's Oxford group,  Mike  discovered with Capper \cite{Capper:1974ic} the Weyl anomaly. The vanishing of the trace of the stress-energy tensor implied by the local scale (Weyl) invariance, first proposed Hermann Weyl in 1918, is not preserved quantum mechanically.   This was a surprise, so much so that it was largely dismissed as wrong \cite{Duff:1993wm} by many of the leading lights of the day\footnote{See also the contribution of Steven M.~Christensen in this same collection.}. Such doubts, however, were quelled by an influential paper of Mike, Stanley Deser and Isham \cite{Deser:1976yx}, which provided the most general form of the trace in various dimensions and made it plain that the anomaly could not be removed by local counterterms. It was there to stay. The possibility of Weyl anomalies is, of course, now universally recognised and has  had tremendous implications across diverse  contexts: Hawking radiation \cite{Christensen:1977jc}; asymptotic safety \cite{Weinberg:1980gg, Christensen:1978sc}; string theory \cite{Polyakov:1981rd}; supersymmetry and supergravity \cite{Ferrara:1974pz,Fradkin:1983tg,Anselmi:1997am, Liu:1998bu}; inflation \cite{Starobinsky:1980te, Hawking:2000bb, Pelinson:2002ef}; holography \cite{Henningson:1998gx, Graham:1999pm}; braneworlds \cite{Karch:2000ct}; condensed matter \cite{Chernodub:2016lbo} and conformal colliders \cite{Hofman:2008ar}. For instance, Tohru Eguchi and Peter G.O. Freund had  identified the Pontryagin number as characterising the    axial fermion number current anomaly, but  noted that there did not seem to be any  analogous role for the Euler characteristic \cite{Eguchi:1976db}. Motivated by  this apparent gap, Mike showed \cite{Duff:1977ay} that the Euler characteristic corresponds to  the integrated trace anomaly, of course! In particular, in $d=2$  dimensions the Weyl anomaly is just $ a R$, where $R$ is the Ricci scalar and $a$ is the anomaly coefficient. In the context of string theory Polyakov famously showed \cite{Polyakov:1981rd} that the vanishing of the world-sheet Weyl anomaly picks out the critical dimensions, where the $a$  anomaly coefficient is related to the Virasoro algebra  central charge by $c=a/24\pi$. Moreover, on including spacetime background fields the vanishing of the world-sheet Weyl anomaly implies the spacetime Einstein equations of (super)gravity \cite{Callan:1985ia, Fradkin:1985fq}, a  remarkable result sitting at the foundations of string theory.

Crossing the pond to Brandeis, Waltham, MA, USA, in 1977, Mike joined forces with  Steven M.~Christensen at Harvard to compute  Weyl and axial  anomalies in the then recently discovered theory of supergravity. In particular,  they were to show that the superpartner to the graviton, the gravitino, contributes an axial anomaly -21 times that of a Dirac spinor \cite{Christensen:1978gi}. This was again met with some disbelief,  but perhaps most interesting was their approach, generalising the classical index theorems, such as Atiyah--Singer, to arbitrary spin  \cite{Christensen:1978md}.  Such calculations revealed some unexpected subtleties. 
Together with Peter van Nieuwenhuizen, Mike demonstrated that the partition function and Weyl anomaly of a given field may not coincide with those of its electromagnetic dual \cite{Duff:1980qv}. Here the anomaly is given by ${\rm tr} \langle T\rangle_{\rm reg}- \langle {\rm tr}  T\rangle_{\rm reg}$, where ${\rm tr} $ denotes the  trace, $\langle - \rangle_{\rm reg}$ is the  regularized expectation value and $T$ is the stress--energy tensor \cite{Duff:1977ay}.  They used this observation to  argue that theories, classically equivalent  under electromagnetic duality, may fail to be so quantum mechanically \cite{Duff:1980qv},  which is by now a well-recognised property of quantum field theory on topologically non-trivial manifolds \cite{Schwarz:1984wk, Donnelly:2016mlc, Kuzenko:2020zad, Borsten:2021pte}.  This anomaly should not be confused with ${\rm tr} \langle T\rangle_{\rm reg}$ alone which yields equivalent results \cite{Siegel:1980ax, Grisaru:1984vk, Bern:2015xsa}.   Fast-forward some 42 years, Mike demonstrated that the Weyl anomaly of (the massless sector of) type IIA string theory compactified on a 6-manifold is given by a product of  Euler characteristics $\chi(M\times X)=\chi(M)\chi(X)$, where $M$ is the (Euclidean) spacetime 4-manifold and $X$ is the internal 6-manifold. Moreover,  for (the massless sector of) M-theory compactified on a 7-manifold $Y$, the Weyl anomaly is given by the product $\rho(M\times Y)=\chi(M)\rho(Y)$, where $\rho(Y)$ is a topological invariant reminiscent of the Ray-Singer torsion \cite{Borsten:2021pte}. If you like, $\rho$ is to M-theory what $\chi$ is to strings.

This early foray into supergravity marked the beginning of Mike's next major movement: Kaluza-Klein theory. In the early 1980s Mike was  to return to Imperial College London and also spend time at CERN, Meyrin, Switzerland, two institutes that played an important role in the development of Kaluza--Klein supergravity.  At this time supergravity offered  much promise as a  unified theory, necessarily including gravity.  First, it was hoped that supersymmetry  might ameliorate the UV divergences plaguing perturbative quantum gravity. Second, supergravity is unique  and particularly elegant in $D=11$ spacetime dimensions, the maximum allowed by supersymmetry. Thus, when combined with Kaluza--Klein compactification, supergravity stood out  as an approach to unification\footnote{So much so, Stephen W.~Hawking dared to title his inaugural lecture as  the Lucasian Professor of Mathematics with the question `Is the End in Sight for Theoretical Physics?',  with $\mathcal{N}=8$ supergravity  in mind. But even Hawking was not exempt from Betteridge's law.}. In this context, Mike and his colleagues made several key advances.  With Christopher N.~Pope, Mike showed that $D=4$, $\text{SO}(8)$ gauged $\mathcal{N}=8$ supergravity theory could be derived as a spontaneous Kaluza-Klein compactification   of $D=11$ supergravity  on $\text{AdS}_4\times S^7$ \cite{Duff:1983gq}. Besides its importance for unification at that time, this particular compactification has been a cornerstone of many of the subsequent advances in supergravity and string/M-theory. With Mike's PhD student, Moustafa A.~Awada, they further showed that by preserving the $S^7$ topology while deforming its geometry one could break the $\mathcal{N}=8$ supersymmetry down to $\mathcal{N}=1$ \cite{Awada:1982pk}. This entailed two important insights that would shape much future work on string/M-theory compactifications. First, the holomony   of the internal manifold dictates the degree of supersymmetry preserved. In the context of heterotic superstring compactifications with vanishing fluxes this famously picks out Calabi--Yau 3-folds as the internal manifolds of choice for model building. Second, Mike, Pope and Bengt E.~W.~Nilsson subsequently  showed that the supersymmetry breaking induced by the squashed $S^7$ corresponded to a Higgs mechanism from the $D=4$ perspective \cite{Duff:1983ajq}. Not long after, the same trio performed the first $K3$ compactification \cite{Duff:1983vj}. This was motivated, in part, by its special $\text{SU}(2)$ holonomy,  a prelude to the all important $\text{SU}(3)$ holonomy Calabi--Yau 3-fold superstring compactifications that would be initiated shortly after \cite{Candelas:1985en}. Moreover, the  $\text{SU}(2)$ holonomy implies that $K3$ compactifications preserve one half of the supersymmetries, opening the door to type IIA on $K3$ and heterotic on $T^4$ string/string dualities. More on that later. These developments, along with manifold  pioneering contributions made by many others (some of whom can be found in this very collection),   were pulled together by Mike, Nilsson and Pope in what has become a standard reference for Kaluza-Klein supergravity \cite{Duff:1986hr}. 

The sharp crescendo of excitement surrounding supergravity   was just as quickly muffled\footnote{For a sense of just how quickly things were moving, during review process of \cite{Duff:1986hr}  a note was added   summarising the developments, which  would divert the attention of most away from $D=11$ supergravity, that had emerged between submission and  acceptance!}.   It had started to seem unlikely that supergravity could ultimately stave off the divergences inherent to a perturbative quantum \emph{field theory} of gravity (almost 50 years on this chapter is still not quite closed, however)  and   Edward Witten had demonstrated  that $D=11$ supergravity compactified on a manifold could not accommodate the chirality needed to make contact with the Standard Model \cite{Witten:1983ux}. 
By the end of 1985 the groundbreaking discoveries of the Green--Schwarz mechanism, heterotic superstrings and Calabi--Yau 3-fold compactifications had firmly, and rightly, cemented themselves as the most promising route to superunification.

Yet, Mike and many like-minded folk had not yet given up on $D=11$. On the one hand, superstrings were not an open and shut case and in his  1987 `Not the standard superstring review' \cite{Duff:1987dk} Mike erred on the side of caution,
\begin{quote}

``In order not to be misunderstood, let me say straight away that I share the conviction that superstrings are the most exciting development in theoretical physics for many years, and that they offer the best promise to date of achieving the twin goals of a consistent quantum gravity and a unification of all the forces and particles of Nature. Where I differ is the degree of emphasis that I would place on the unresolved problems
of superstrings, and the likely time scales involved before superstrings (or something like superstrings) make contact with experimental reality.''
\end{quote}
He emphasised, in particular, the challenges (and opportunities) posed by the landscape problem and non-perturbative phenomena, such as black holes. On the other hand, the tension between 10 and 11 raised its own questions. Why did supersymmetry allow for 11, while superstrings only 10? If supergravity was the low-energy effective field theory of superstrings, where did that leave $D=11$ supergravity? Mike vigorously maintained that 11 should be taken seriously.   

Indeed, various clues that $D=11$ might yet play a role had been amassing. While there are no superstrings in $D=11$, there are supermembranes that couple to $D=11$ supergravity \cite{Bergshoeff:1987cm}.  It turns out that this is one of the key bridges between $D=10$ string theory and $D=11$ M-theory. In 1987 Mike, Paul S.~Howe, Takeo Inami and Kellogg Stelle showed \cite{Duff:1987bx} by compactifying the $D=11$ spacetime manifold on  $S^1$  and simultaneously wrapping the supermembrane  around  the circle  ones finds precisely the type IIA superstring in $D=10$! This result pre-empted\footnote{In his review of string theory \cite{Conlon:2016kat}, Joseph Conlon remarks that  ``When I first read this paper I was quite shocked by its existence; according to the supposed history of string theory that I had `learned', such a paper could not have been written for almost another decade.''} important facets of the M-theory revolution of 1995 by connecting strings and membranes, along with 10 and 11 dimensions. In the same year, Mike and Miles P.~Blencowe, again inspired by the discovery of supermembranes in $D=11$,    conjectured the existence of super  $p$-branes on the $S^1\times S^p$ boundary of $\text{AdS}_{p+2}$ and presented the corresponding (free) superconformal field theories \cite{Blencowe:1987bn}. The maximal $p=2$ case corresponded to the supermembrane on $\text{AdS}_{4}\times S^7$ with  the superconformal group $\text{OSp}(8|4)$.  The maximal $p=3$ and $p=5$ cases corresponded to the yet to be discovered D3-brane and M5-brane on $\text{AdS}_{5}\times S^5$ and $\text{AdS}_{7}\times S^4$ with  superconformal groups $\text{SU}(2,2|5)$ and  $\text{OSp}(8^*|4)$, respectively.

Further telling clues on the road to M-theory arose in the context of branes and dualities, themselves closely related. By the mid-eighties  five \emph{a priori} independent consistent superstring theories had been established. However, they were not islands; for instance, the IIA and IIB theories   could be connected by T-duality or mirror symmetry.  What was to emerge  over the next decade or so was a web of dualities, suggesting that each string theory was but a corner of a larger framework. During this period of intense activity (in 10 and 11 dimensions) Mike relocated  to Texas A\&M, just in time for it to host the inaugural `Strings 89' conference. Aptly, that year Mike addressed the question of \emph{manifest} T-duality  \cite{Duff:1989mp}.  By considering two \emph{dual} string theories, he introduced the notion of a doubled spacetime  with a generalised $\text{O}(D,D)$ metric $H(g, B)$, built from the standard metric $g$ and the Kalb-Ramond two-form $B$. This is, today, a key ingredient in the thriving domain of double field theory. The following year `Strings' would return to Texas A\&M and this time around Mike and his then PhD student, Jian Xin Lu, generalised these notions to membranes, where the role $B$ is replaced by the three-form $C$ of $D=11$ supergravity \cite{Duff:1990hn}. The goal here was to make manifest, from the membrane's perspective, the global symmetries of  $D=11$ supergravity compactified on an $n$-torus, which would later be recognised as shadows of the U-dualities of M-theory.  This time the spacetime is not merely doubled, but extended by $C^{n}_{2}$ coordinates corresponding to the possible ways one can wrap a membrane on an $n$-torus. There is a generalised metric $H(g, C)$ manifesting the appropriate symmetry group; for example, $\text{SL}(5, \mathbb{R})$ for $n=4$. It is interesting to note that Mike and Lu puzzled over the cases $n>4$, which does not na\"ively work out as expected. They resolved this question, quite naturally, by introducing additional coordinates corresponding to the Hodge dual of $C$ and so recovered the symmetries of $D=11$ supergravity on an $n$-torus, for $1\leq n\leq 8$. Of course, we now understand these coordinates as corresponding to the  possible wrappings of the M5-brane that kick in at $n=5$. The extended spacetimes and their generalised metrics $H(g, C)$ are, today,  central to the developments of exceptional field theory.

Another central theme of Mike's time as a Texan was the role of solitonic supersymmetric $p$-brane solutions that carry topological magnetic charge, and their dual relationship to elementary singular $(D-p-4)$-brane solutions carrying electric Noether charge \cite{Duff:1990wv,Duff:1991pea, Duff:1993ye, Duff:1994fg}. For example, in 1991 Mike and Stelle \cite{Duff:1990xz} discovered the elementary multiple membrane solutions of $D=11$ supergravity,  shortly followed  by the dual solitonic superfivebrane solution of Gueven \cite{Gueven:1992hh}. An other idea introduced by Mike,  with Ramzi R.~Khuri,  Ruben Minasian,  and Joachim Rahmfeld,  during this period was the identification of solitonic magnetic string states as extremal black holes \cite{Duff:1993yb}. Then applying S-duality led Mike and Rahmfeld to relate supersymmetric massive string states with elementary black holes  \cite{Duff:1994jr}. These ideas generalise to the black and super $p$-branes solutions in various dimensions \cite{Duff:1993ye} and have become a key concept in the understanding of black holes in string/M-theory. The profound contributions unravelling  this web of ideas, by Mike and many others, are far too numerous to do justice to here. Fortunately, Mike, Lu and Khuri put together an influential review \cite{Duff:1994an} of these developments up to 1994 that we can defer to. An important consequence of the $p/(D-p-4)$-brane  dualities  \cite{Duff:1995wd} is the implied equivalences  among string compactifications; for example,   the $D = 10$ heterotic string compactified on a 4-torus is quantum  equivalent  to the $D = 10$ type IIA string on $K3$. 

Another related idea introduced by Mike and his colleagues at Texas A\&M, including Lu,  Khuri, Minasian as well as the  newer arrival James T. Liu, was that $p$-brane dualities could be used to explain electromagnetic duality in lower dimensions \cite{Duff:1993ij}, as described by  Witten \cite{Witten:2014int}:
\begin{quote}
``Mike Duff and Ramzi Khuri in 1993 had written a paper on what they called string/string duality. They had said there should be a self-dual string theory in six dimensions that, looked at in two different ways, would give electric-magnetic duality of gauge theory in four dimensions. It was actually a brilliant idea. The only trouble was they didn't have an example in which it worked.''
\end{quote}
Mike and his colleagues rapidly developed an intricate web of dualities amongst strings and $p$-branes, and their implications for strong/weak coupling dualities, over the following years  \cite{Duff:1994zt, Duff:1995sm, Duff:1995sm, Duff:1994zt}. Note,  the particular case of the self-dual string in $D=6$ relates to the role of the $(2,0)$ theory in the geometric Langlands programme. In particular, Mike, Liu and Minasian gave evidence  that  membrane/fivebrane duality provides an eleven  dimensional origin of string/string duality, which in turn bolsters the S-duality conjecture \cite{Duff:1995wd}. The original hope of Mike and Khuri was also realised together with Minasian and Witten in the context of a heterotic/heterotic duality \cite{Duff:1996rs}.   These observations contributed (along with   crucial insights of a great many others that we are shamefully unable to pay due homage to here) to the 1995 M-theory revolution led by Witten, which marked a new phase in the development of strings and branes. The supermembrane and fivebrane were duly promoted to the M2- and M5-brane and $D=11$ found its place, after all,  as the low-energy limit of M-theory. Mike's conviction that $D=11$ should be   canon was vindicated. 

\begin{figure}[h]
    \centering
    \includegraphics[width=0.233\textwidth]{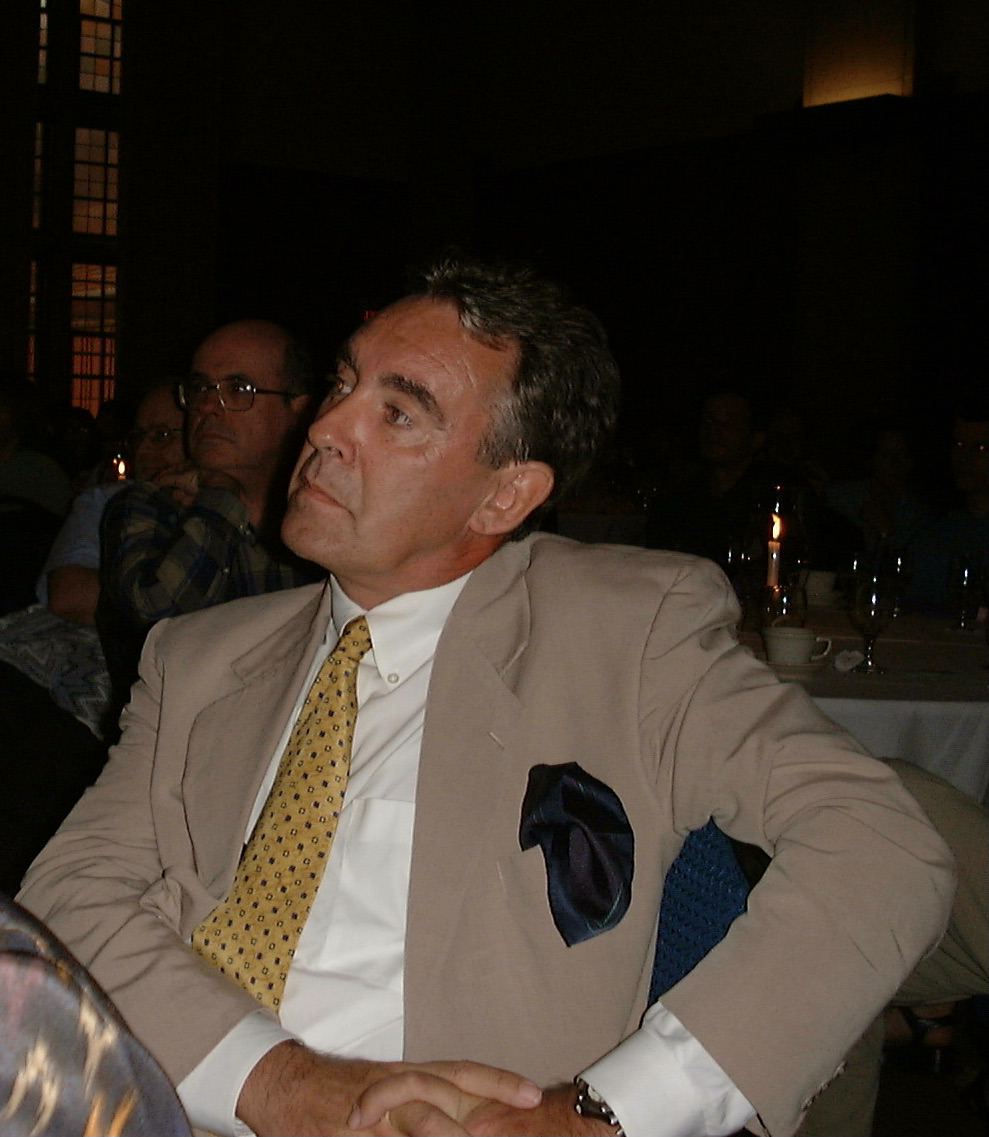}    \includegraphics[width=0.356\textwidth]{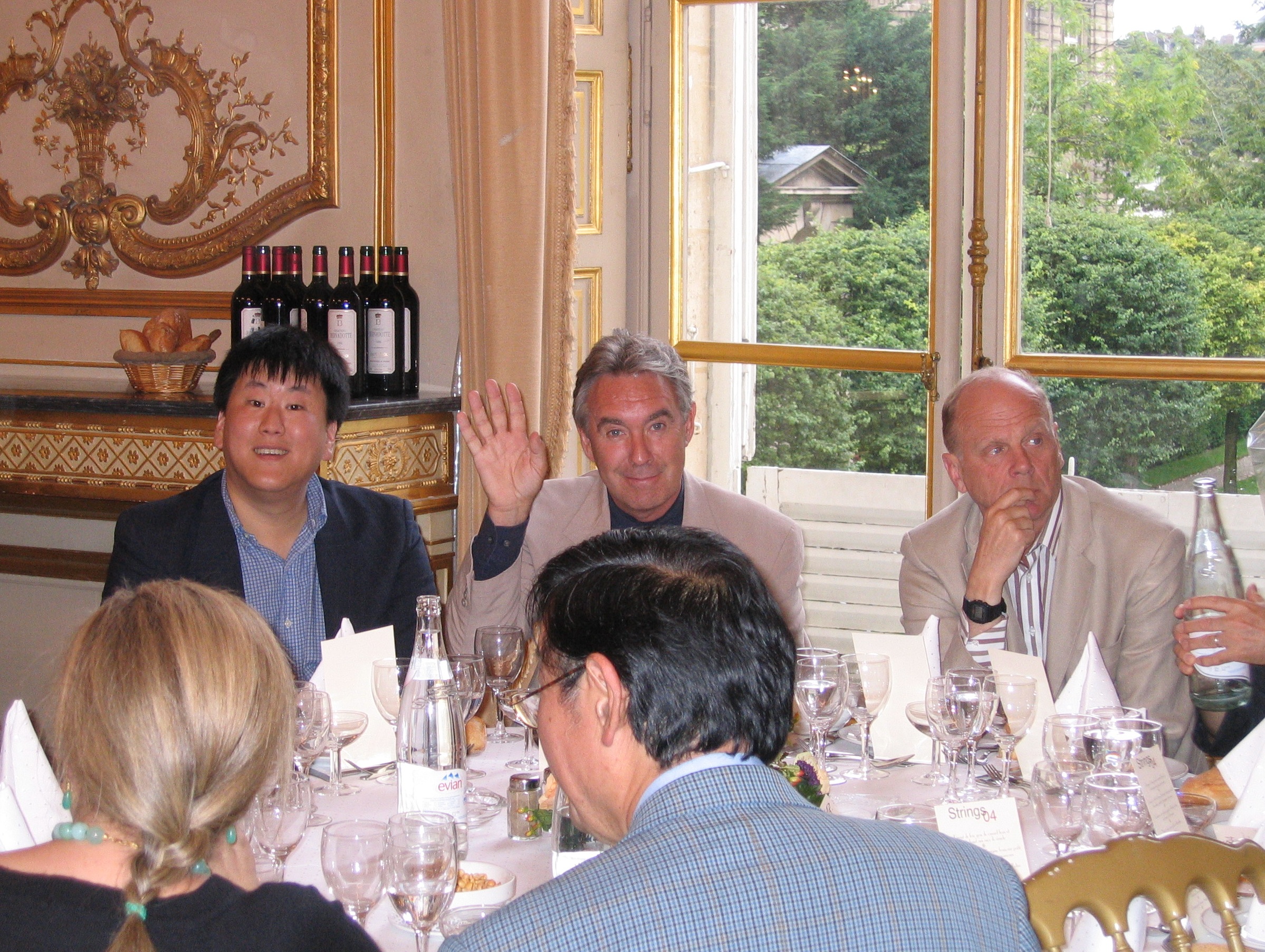}
    \includegraphics[width=0.399\textwidth]{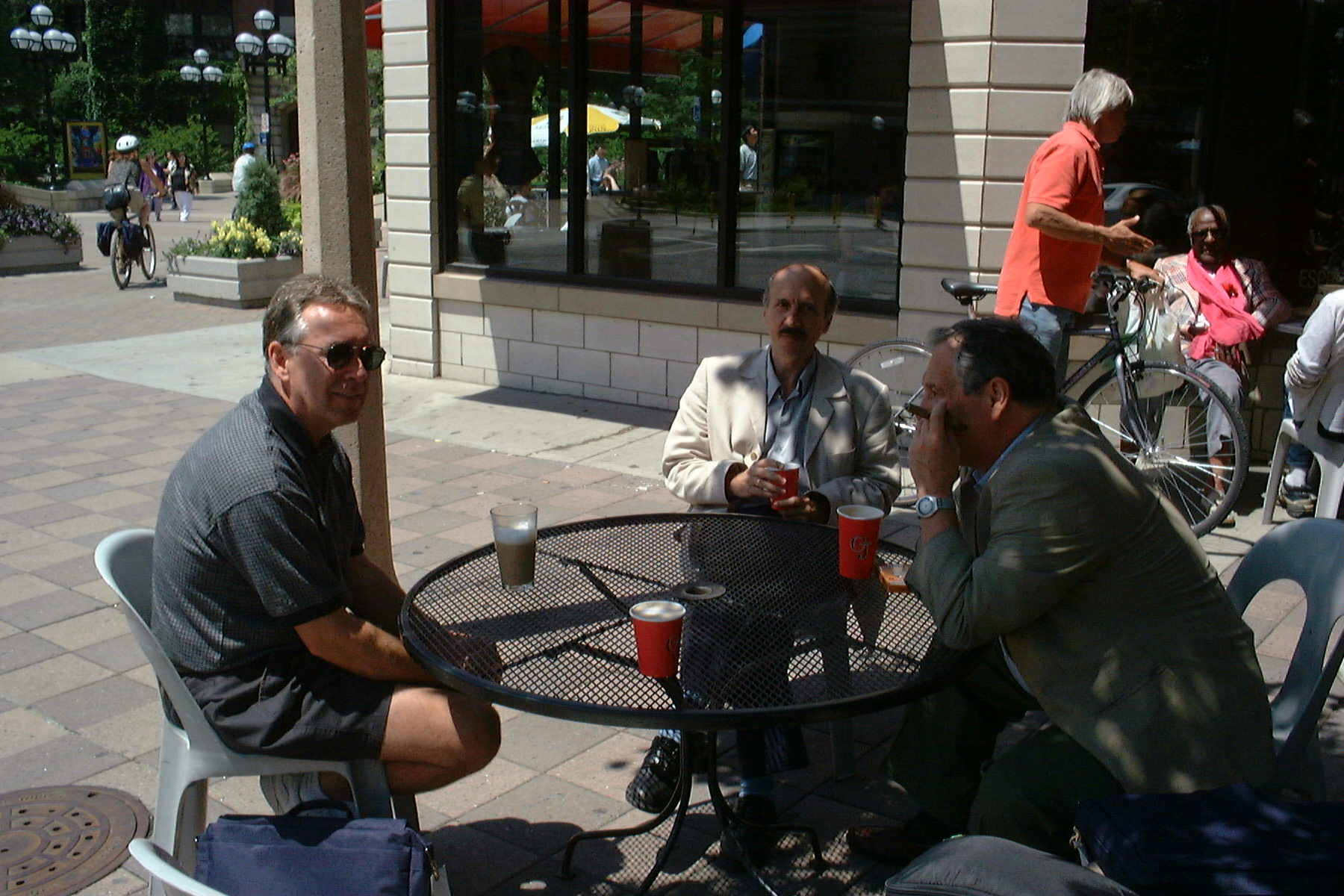}
    \caption{Mike presiding over  `Strings 2000' at the University of Michigan, with (left to right) James T.~Liu, Lars Brink, Ignatios Antoniadis and Sergio Ferrara.}
    \label{fig:2000}
\end{figure}

This period was followed by an explosion of ideas in string/M-theory, the anti-de Sitter/conformal field theory correspondence (AdS/CFT) and Randall--Sundrum brane world models to name but two examples. Mike's past work,  such as the AdS compactifications and brane-scans \cite{Duff:1992hu}, fed into many aspects of the renewed research avenues. In fact, jumping ahead a little, his 1973 work on loop corrections to the Schwarzschild solution proved important to (AdS/CFT)/Randall--Sundrum complementarity, as shown by Mike and Liu  \cite{Duff:2000mt}. Who in 1973 saw that one coming!
 In particular, Mike and colleagues developed asymptotically flat and AdS black hole and $p$-brane solutions to M-theory \cite{Duff:1996hp, Duff:1999gh, Duff:1998us, Cvetic:1999xp}, crucial to the applications of AdS/CFT   and  the question of Bekenstein-Hawking entropy.  During this period Mike  left the Texas triangle to, fittingly, take up the Oskar Klein professorship at the University of Michigan, where he would be elected as the first director of the newly created Michigan Center for Theoretical Physics.  Again, Mike arrived just in time for Michigan to host Strings, this time the millennial edition,  `Strings 2000' (\autoref{fig:2000}).
 
 To a degree it was time to take stock. Mike, David Gross and Witten solicited `big questions' from the attendees and selected the ten best. Some transcended any particular approach to physics beyond the standard model or quantum gravity, for example `Why does the universe appear to have one time and three space dimensions?'. But one was squarely in the domain of M-theory: 
 \begin{quote}
 ``What are the fundamental degrees of freedom of M-theory (the theory whose low-energy limit is eleven-dimensional supergravity and which subsumes the five consistent superstring theories) and does the theory describe Nature?''  
 \end{quote}
 This is perhaps the question with which  Mike himself has since been most preoccupied: elucidating what M-theory \emph{is}.   Although we have collectively uncovered a patchwork understanding, its ultimate formulation requires new ideas and insights, an endeavour  Mike has constantly championed.

In 2005 Mike would come full circle, returning to Imperial College London, now as the Abdus Salam Professor of Theoretical Physics. Here he would embark on several new research journeys (but always touching on M-theory), such as black holes and qubits \cite{Borsten:2012fx}, quantum optics and Hawking radiation \cite{Ben-Benjamin:2019opz} and gravity as the `square' of Yang--Mills theory \cite{Borsten:2015pla}. 

For instance, not long after arriving, Mike and  Sergio Ferrara,  with various of  their colleagues and students, would build a dictionary between string/M-theory  black holes and various concepts from quantum information theory, qubits and entanglement measures  \cite{Duff:2006uz, Duff:2006ue, Duff:2007wa, Borsten:2008ur, Borsten:2008wd, Borsten:2010db, Borsten:2012fx}. This programme grew out of the observation that the entropy of the $STU$ black hole\footnote{$STU$ supergravity was  introduced by Mike, Liu and Rahmfeld back in Texas. It is  special in that its symmetries  correspond not to a string/string duality, but a string/string/string triality!   It has since been a paradigmatic model  for elucidating facets of string/M-theory. It also connects black hole entropy to the number theory of Manjul Bhargava \cite{Borsten:2020nqi}.} and the entanglement shared by three qubits are both described by Cayley's hyperdeterminant \cite{Duff:2006uz}. One can only assume that Cayley had anticipated  both M-theory and quantum computing. 

In completely separate developments, Mike initiated a programme to understand `Einstein as the square of Yang--Mills' at the level of off-shell field theories. The notion of gravity as the `product' of two gauge theories has a long history, but was in particular made concrete through the tree-level Kawai--Lewellen--Tye  `closed $=$ open $\times$ open' string scattering relations. This idea has witnessed  a recent renaissance driven by the  2008 Bern-Carrasco-Johansson colour/kinematics duality conjecture, which allows one to build graviton scattering amplitudes from the `double copy' of gluon amplitudes to all orders in perturbation theory.\footnote{This  reopened the debate concerning the perturbative finiteness of $\mathcal{N}=8$ supergravity, costing one of us some bottles of wine.}  Inspired, in part,  by the relationship between the symmetries of supergravity and those of  super Yang-Mills theory, Mike took an off-shell  field theory approach to `gravity $=$ gauge $\times$ gauge'. This  yielded remarkable and unexpected insights such as the appearance of the Freudenthal magic square of U-dualities \cite{Borsten:2013bp, Anastasiou:2013hba} and the Yang-Mills origin of  (super)diffeomorphisms \cite{Anastasiou:2014qba, Anastasiou:2018rdx}. Today  scattering amplitudes and the `double copy' are  being used to understand classical gravity, in particular black hole collisions. We would imagine that  Salam (or a 20-year-old Mike)  would have been  surprised and delighted at this.

\ack{We are most grateful to all the authors for their wonderful contributions to this collection. We are also immensely  grateful to the Proceedings of the Royal Society Publishing Editor Joanna Harries, and all the  PRSA staff, for her tremendous work and support in bringing this special issue together.}


\end{document}